\documentclass[12pt]{amsart}

\usepackage{amsfonts}

\newtheorem{theorem}{Theorem}[section]
\theoremstyle{plain}

\newtheorem{condition}{Condition}

\newtheorem{definition}[theorem]{Definition}

\newtheorem{lemma}[theorem]{Lemma}

\newtheorem{proposition}[theorem]{Proposition}

\numberwithin{equation}{section}
\input{tcilatex}

\begin{document}
\title[Quantum Sequential Product]{Characterization of the Sequential
Product on Quantum Effects}
\author{Stan Gudder}
\address{Department of Mathematics\\
University of Denver\\
Denver CO 80208}
\email{sgudder@math.du.edu}
\urladdr{http://www.math.du.edu/\symbol{126}sgudder}
\author{Fr\'{e}d\'{e}ric Latr\'{e}moli\`{e}re}
\address{Department of Mathematics\\
University of Denver\\
Denver CO 80208}
\email{frederic@math.du.edu}
\urladdr{http://www.math.du.edu/\symbol{126}frederic}
\date{1/10/2008}

\begin{abstract}
We present a characterization of the standard sequential product of quantum
effects. The characterization is in term of algebraic, continuity and
duality conditions that can be physically motivated.
\end{abstract}

\maketitle

\section{Introduction}

This paper gives a set of five physically motivated conditions which fully
characterize the sequential product on quantum effects. The positive
operators on a complex Hilbert space $\mathcal{H}$ that are bounded above by
the identity operator $I$ are called the \emph{quantum effects} on $\mathcal{
\ H}$. The set of quantum effects on $\mathcal{H}$ is denoted by $\mathcal{E}
\left( \mathcal{H}\right) $. Quantum effects represent yes-no measurements
that may be unsharp. The subset $\mathcal{P}\left( \mathcal{H}\right) $ of $
\mathcal{E}\left( \mathcal{H}\right) $ consisting of orthogonal projections
represent sharp yes-no measurements. Another important subset of $\mathcal{E}
\left( \mathcal{H}\right) $ is the set $\mathcal{D}\left( \mathcal{H}\right) 
$ of density operators, i.e. the trace-class operators on $\mathcal{H}$ of
unit trace, which represent the states of quantum systems. If $A\in \mathcal{
\ E}\left( \mathcal{H}\right) $ and $\rho \in \mathcal{D}\left( \mathcal{H}
\right) $ then $\limfunc{Tr}(\rho A)$ is the probability that $A$ is
observed (the answer is yes) when the system is in the state $\rho $.

\bigskip A \emph{sequential product }defined by $A\circ B=A^{\frac{1}{2}
}BA^{ \frac{1}{2}}$ for any two quantum effects $A,B$ has recently been
introduced and studied \cite{AriasGudder, Gheandera04, Gudder02, Gudder05,
Gudder01, Leifer06, Molnar03}. The product $A\circ B$ represents the effect
produced by first measuring $A$ then measuring $B$. This product has also
been generalized to an algebraic structure called a \emph{sequential effect
algebra (SEA)}. Examples of SEA are $\left[ 0,1\right] \subseteq \mathbb{R}$
, Boolean algebras, fuzzy set systems $\left[ 0,1\right] ^{X}$ and $\mathcal{
\ E}\left( \mathcal{H}\right) $ . It has been shown that the sequential
product is unique on all of these structures except $\mathcal{E}\left( 
\mathcal{H}\right) $ and it has been an open problem whether $A\circ B=A^{ 
\frac{1}{2}}BA^{\frac{1}{2}}$ is the unique sequential product on $\mathcal{
E }\left( \mathcal{H}\right) $. It would be important physically to
establish this uniqueness because we would then have an unambiguous form for
the quantum mechanical sequential product.

\bigskip There are various reasons for the appeal of the form $A\circ B=A^{ 
\frac{1}{2}}BA^{\frac{1}{2}}$ ($A,B\in \mathcal{E}\left( \mathcal{H}\right) $
). First, when $P$ and $Q$ are orthogonal projections, then $P\circ Q=PQP$
is the accepted form for an ideal measurement in that case \cite{Busch95,
Busch96, Davis76}. Second, $\circ $ satisfies various algebraic, continuity
and duality conditions that one would expect from a sequential product. For
example, for all $A,B,C\in \mathcal{E}\left( \mathcal{H}\right) $ we have $
I\circ A=A\circ I=A$, as well as $A\circ \left( B+C\right) =A\circ B+A\circ
C $ whenever $B+C\in \mathcal{E}\left( \mathcal{H}\right) $, $A\circ B\leq A$
and for all $\lambda \in \left[ 0,1\right] $ we have $\lambda \left( A\circ
B\right) =\left( \lambda A\right) \circ B=A\circ \left( \lambda B\right) $.
Moreover, $\circ $ is jointly continuous for the strong operator topology.
Finally, for any state $\rho \in \mathcal{D}\left( \mathcal{H}\right) $ and
quantum effects $A,B\in \mathcal{E}\left( \mathcal{H}\right) $ we have the
duality relation $\limfunc{Tr}\left( \rho (A\circ B\right) )=\limfunc{Tr}
\left( \left( A\circ \rho \right) B\right) $. We shall discuss the physical
motivations for these conditions in the next section of this paper.

Our last reason for accepting the form $A\circ B=A^{\frac{1}{2}}BA^{\frac{1}{
2}}$ stems from quantum computation and information theory \cite{Nielsen00}.
If $\left( A_{i}\right) _{i\in \mathbb{N}}$ is a sequence of bounded linear
operators on $\mathcal{H}$ satisfying $\sum_{i=0}^{\infty }A_{i}^{\ast
}A_{i}=I$ then the operators $A_{i}$ ($i\in \mathbb{N}$) are called the 
\emph{operational elements }of the \emph{quantum operation} $\mathfrak{A}: 
\mathcal{D}(\mathcal{H})\longrightarrow \mathcal{D}(\mathcal{H})$ defined
by: 
\begin{equation}
\mathfrak{A}\left( \rho \right) =\sum_{i=0}^{\infty }A_{i}\rho A_{i}^{\ast } 
\text{.}  \label{QuantumOp}
\end{equation}
Technically speaking, any trace preserving, normal, completely positive map
has the form (\ref{QuantumOp}). Quantum operations are ubiquitous in quantum
computation and information theory. They are used to describe dynamics,
measurements, quantum channels, quantum interactions and quantum error
correcting codes.

For a quantum measurement with outcomes labeled by $\mathbb{N}$, the
operator $\mathfrak{A}\left( \rho \right) $ is the output state produced
after the measurement is performed with input $\rho \in \mathcal{D}\left( 
\mathcal{H}\right) $. If the outcome $i\in \mathbb{N}$ occurs, then an axiom
of quantum mechanics says that the post measurement state becomes: 
\begin{equation}
\left( \rho |A_{i}\right) =\frac{A_{i}\rho A_{i}^{\ast }}{\limfunc{Tr}\left(
A_{i}\rho A_{i}^{\ast }\right) }\text{.}  \label{QuantumStateChange}
\end{equation}

Now, a very general type of measurement is called an \emph{observable} and
is modeled by a \emph{positive operator-valued measure }(POVM).\ To keep
this discussion simple, we shall only consider discrete observables. In this
case, we can label the outcomes as before by $\mathbb{N}$ and the effect
that the observable has outcome $i\in \mathbb{N}$ is denoted by $E_{i}\in 
\mathcal{E}\left( \mathcal{H}\right) $. Since one of the outcomes is always
observed, we have $\sum_{i=0}^{\infty }E_{i}=I$. Therefore $
\sum_{i=0}^{\infty }\left( E_{i}^{\frac{1}{2}}\right) ^{\ast }\left( E_{i}^{ 
\frac{1}{2}}\right) =\sum_{i=0}^{\infty }E_{i}=I$ so $\left( E_{i}^{\frac{1}{
2}}\right) _{i\in \mathbb{N}}$ is the sequence of operational elements for
the quantum operation $\mathfrak{A}:\rho \in \mathcal{D}\left( \mathcal{H}
\right) \mapsto \sum_{i=0}^{\infty }E_{i}^{\frac{1}{2}}\rho E_{i}^{\frac{1}{
2 }}$, and (\ref{QuantumStateChange}) becomes: 
\begin{equation}
\left( \rho |E_{i}^{\frac{1}{2}}\right) =\frac{E_{i}^{\frac{1}{2}}\rho
E_{i}^{\frac{1}{2}}}{\limfunc{Tr}\left( E_{i}^{\frac{1}{2}}\rho E_{i}^{\frac{
1}{2}}\right) }\text{.}  \label{QuantumStateChange2}
\end{equation}

Now, the real number $\mathbb{P}_{\rho }\left( E_{i}\right) $ defined by $
\mathbb{P}_{\rho }\left( E_{i}\right) =\limfunc{Tr}\left( \rho E_{i}\right) $
is the probability that outcomes $i\in \mathbb{N}$ occurs in the state $\rho 
$ and we can write (\ref{QuantumStateChange2}) as: 
\begin{equation}
E_{i}^{\frac{1}{2}}\rho E_{i}^{\frac{1}{2}}=\mathbb{P}_{\rho }\left(
E_{i}\right) \left( \rho |E_{i}\right) \text{.}  \label{Proba1}
\end{equation}
We can extend the quantum operation $\mathfrak{A}$ to $\mathcal{E}\left( 
\mathcal{H}\right) $ and thus obtain, for all $F\in \mathcal{E}\left( 
\mathcal{H}\right) $: 
\begin{equation}
E_{i}^{\frac{1}{2}}FE_{i}^{\frac{1}{2}}=\mathbb{P}_{\rho }(E_{i})\left(
F|E_{i}\right) \text{.}  \label{Proba2}
\end{equation}
Now, (\ref{Proba2}) is formally analogous to the formula for conditional
probability in classical probability theory. In that case, it seems
reasonable to interpret the left hand side of (\ref{Proba2}) as the formula
for \textquotedblleft $E_{i}$ and $F$\textquotedblright . However, $
(F|E_{i}) $ is not symmetric in $F$ and $E_{i}$ but rather supposes that $
E_{i}$ was measured first. In the present noncommutative setting, we more
precisely interpret $E^{\frac{1}{2}}FE^{\frac{1}{2}}$ ($E,F\in \mathcal{E}
\left( \mathcal{H}\right) $) as the effect obtained from measuring $E$ first
and $F$ second.

\section{Physical Motivations}

This section gives physical motivations for conditions that we shall use to
characterize the sequential product on quantum effects. From now on in this
paper, we shall always use $\circ $ to designate a general product on $
\mathcal{E}\left( \mathcal{H}\right) $ which satisfies the conditions given
in this section. Later we shall establish that for all $A,B\in \mathcal{E}
\left( \mathcal{H}\right) $ we have $A\circ B=A^{\frac{1}{2}}BA^{\frac{1}{2}
} $ and thus that our conditions uniquely determine the sequential product
on quantum effects.

\bigskip A sequential product has two dual roles. When $A$ and $B$ are
quantum effects, then $A\circ B$ is itself a quantum effect whose physical
interpretation should be the effect measuring $B$ after measuring $A$. On
the other hand, given a state $\rho \in \mathcal{D}\left( \mathcal{H}\right) 
$, since then $\rho \in \mathcal{E}\left( \mathcal{H}\right) $ we can form $
A\circ \rho $ for all $A\in \mathcal{E}$ $\left( \mathcal{H}\right) $. We
shall impose on $\circ $ that the relation $\limfunc{Tr}(A\circ \rho )= 
\limfunc{Tr}(\rho A)$ must hold for all $A\in \mathcal{E}\left( \mathcal{H}
\right) $ --- though in fact we will eventually retain a more general
condition. In other words, $A\circ \rho $ is a trace-class operator whose
trace is the probability $\mathbb{P}_{\rho }(A)$ of observing $A$ in $\rho $
. From this, given any effect $B$, it is natural to interpret the
probability $\limfunc{Tr}(\left( A\circ \rho \right) B)$ as the probability
to observe $B$ and $A$ in the state $\rho $, with the additional assumption
that $A$ is measured first. Let us assume now that $\mathbb{P}_{\rho }(A)= 
\limfunc{Tr}(A\circ \rho )\not=0$. Then, it is natural to define the
conditional probability of observing $B$ given that $A$ is observed first in
the state $\rho $ as the probability $\mathbb{P}_{\rho |A}(B)$ defined by: 
\begin{equation*}
\mathbb{P}_{\rho |A}(B)=\frac{\limfunc{Tr}(\left( A\circ \rho \right) B)}{ 
\limfunc{Tr}(A\circ \rho )}\text{.}
\end{equation*}

\bigskip On the other hand, the probability of $B$ given that $A$ is
observed first, computed in the original state $\rho $, should be given by: 
\begin{equation*}
\mathbb{P}_{\rho }(B|A)=\frac{\limfunc{Tr}(\rho (A\circ B))}{\limfunc{Tr}
\left( A\circ \rho \right) }
\end{equation*}
since $A\circ B$ precisely represents the effect of observing $B$ after $A$.
It appears reasonable to impose on $\circ $ that both these probabilities
should be equal as they should describe the same event. Thus, if $\limfunc{
Tr }(A\circ \rho )\not=0$ we should have $\mathbb{P}_{\rho |A}(B)=\mathbb{P}
_{\rho }(B|A)$. Simplifying by $\limfunc{Tr}(A\circ \rho )$ and generalizing
to all of $\mathcal{E}\left( \mathcal{H}\right) $ gives us:

\begin{condition}
\label{Duality}\emph{(Duality) }A sequential product $\circ $ satisfies the
relation: 
\begin{equation*}
\limfunc{Tr}(\left( A\circ \rho \right) B)=\limfunc{Tr}(\rho (A\circ B))
\end{equation*}
for all states $\rho \in \mathcal{D}\left( \mathcal{H}\right) $ and all
quantum effects $A,B\in \mathcal{E}\left( \mathcal{H}\right) $.
\end{condition}

\bigskip Note that since $\rho \in \mathcal{D}\left( \mathcal{H}\right) $ is
trace-class, so is $\rho (A\circ I)$ and thus Condition (\ref{Duality})
implies that $\limfunc{Tr}\left( A\circ \rho \right) =\limfunc{Tr}\left(
\rho (A\circ I)\right) $ so $A\circ \rho $ is trace-class as well, of trace
in $\left[ 0,1\right] $. Now, Condition (\ref{Duality}) implies that $\circ $
must be affine in its second variable. It will be useful to record this fact
for our discussion:

\begin{lemma}
\label{Convexity2}Let us assume that $\circ $ satisfies Condition (\ref
{Duality}). Then for all $A,B,C\in \mathcal{E}\left( \mathcal{H}\right) $
and all $\lambda \in \left[ 0,1\right] $ we have: 
\begin{equation*}
A\circ \left( \lambda B+\left( 1-\lambda \right) C\right) =\lambda \left(
A\circ B\right) +\left( 1-\lambda \right) \left( A\circ C\right) 
\end{equation*}
i.e. $B\mapsto A\circ B$ is affine on the convex set $\mathcal{E}\left( 
\mathcal{\ H}\right) $.

In particular, if $\eta $ is a trace class operator on $\mathcal{H}$ with
trace $\lambda \in \left[ 0,1\right] $ then for all $A,B\in \mathcal{E}
\left( \mathcal{H}\right) $ we have: 
\begin{equation*}
\limfunc{Tr}\left( \left( A\circ \eta \right) B\right) =\limfunc{Tr}\left(
\eta \left( A\circ B\right) \right) \text{.}
\end{equation*}
\end{lemma}

\begin{proof}
Let $A,B,C\in \mathcal{E}\left( \mathcal{H}\right) $ and $\rho \in \mathcal{
\ D }\left( \mathcal{H}\right) $ for all this proof.

Let $\lambda \in \left[ 0,1\right] $. Then: 
\begin{eqnarray*}
\limfunc{Tr}\left( \rho \left( A\circ \left( \lambda B\right) \right)
\right) &=&\limfunc{Tr}\left( \left( A\circ \rho \right) \left( \lambda
B\right) \right) \text{ by Condition (\ref{Duality}),} \\
&=&\lambda \limfunc{Tr}\left( \left( A\circ \rho \right) B\right) \\
&=&\lambda \limfunc{Tr}\left( \rho \left( A\circ B\right) \right) \text{ by
Condition (\ref{Duality}) again.}
\end{eqnarray*}
Since $\rho $ is arbitrary, we deduce that $A\circ \left( \lambda B\right)
=\lambda \left( A\circ B\right) $.

Thus, let $\eta $ be a trace-class operator of trace $\lambda \in (0,1]$.
Then: 
\begin{eqnarray*}
\limfunc{Tr}\left( \eta \left( A\circ B\right) \right) &=&\lambda \limfunc{
Tr }\left( \frac{1}{\lambda }\eta \left( A\circ B\right) \right) \\
&=&\lambda \limfunc{Tr}\left( \left( A\circ \left( \frac{1}{\lambda }\eta
\right) \right) B\right) \text{ by Condition (\ref{Duality}),} \\
&=&\limfunc{Tr}\left( \left( A\circ \eta \right) B\right) \text{ by our work
above.}
\end{eqnarray*}

We prove additivity in a similar manner. We have: 
\begin{eqnarray*}
\limfunc{Tr}\left( \rho \left( A\circ \left( B+C\right) \right) \right) &=& 
\limfunc{Tr}\left( \left( A\circ \rho \right) \left( B+C\right) \right) 
\text{ by Condition (\ref{Duality}),} \\
&=&\limfunc{Tr}\left( \left( A\circ \rho \right) B\right) +\limfunc{Tr}
\left( \left( A\circ \rho \right) C\right) \\
&=&\limfunc{Tr}\left( \rho \left( A\circ B+A\circ C\right) \right)
\end{eqnarray*}
where we used Condition (\ref{Duality}) again. Once again, since $\rho $ is
an arbitrary state, we conclude that $A\circ \left( B+C\right) =A\circ
B+A\circ C$.
\end{proof}

\bigskip The identity $I$ of $\mathcal{H}$ is the effect which always
measures $1$, or yes, no matter what state the quantum system is in.
Consequently, measuring $I$ does not affect the quantum system (which
reflects the fact that $I$ commutes with all operators). So measuring $I$
before or after measuring $A\in \mathcal{E}\left( \mathcal{H}\right) $
should not change the simple measurement of $A$. Formally, we shall
henceforth assume that:

\begin{condition}
\label{Unital}\emph{(Unit)} A sequential product $\circ $ needs to satisfy: 
\begin{equation*}
A\circ I=I\circ A=A
\end{equation*}
for all $A\in \mathcal{E}\left( \mathcal{H}\right) $.
\end{condition}

\bigskip We note that given $\rho \in \mathcal{E}\left( \mathcal{H}\right) $
and $A,B\in \mathcal{E}\left( \mathcal{H}\right) $ we have: 
\begin{eqnarray}
\limfunc{Tr}\left( \left( A\circ \rho \right) B\right) &=&\limfunc{Tr}\left(
\left( A\circ \rho \right) \left( B\circ I\right) \right) \text{ by
Condition (\ref{Unital}),}  \notag \\
&=&\limfunc{Tr}\left( B\circ \left( A\circ \rho \right) \right) \text{ by
Lemma\ (\ref{Convexity2}),}  \label{halfDuality}
\end{eqnarray}
since $A\circ \rho $ is trace-class of trace in $\left[ 0,1\right] $.

\bigskip More generally, suppose we are given two quantum effects $A,B$. Let
us assume that $A$ and $B$ commute. Physically, we are therefore assuming
that measurements of $A$ do not affect $B$ and vice-versa. Therefore, the
sequential product should be symmetric: measuring $A$ first and $B$ second
should be the same as measuring $B$ first and $A$ second. Even more
concretely, since $A$ and $B$ commute, they can be measure simultaneously
and this measurement is given by the effect $AB$. Thus, we can physically
expect that $A\circ B=AB=BA=B\circ A$. We actually will only require a
special case of this observation: namely, that $A^{2}=A\circ A$ for all $
A\in \mathcal{E}\left( \mathcal{H}\right) $.

Let us generalize this principle further. Let $\rho \in \mathcal{D}\left( 
\mathcal{H}\right) $ be a state of a quantum system. Let $A,B$ be two
quantum effects. We can view $A$ and $B$ as two successive unsharp filters.
We can proceed with a first experiment by sending the state $\rho $ through $
A$ and measure the resulting state as $\frac{1}{\limfunc{Tr}\left( A\circ
\rho \right) }\left( A\circ \rho \right) $. Thus our quantum system is in a
new state, and can be sent through the second filter $B$. Measuring the
state at the exit of $B$ we shall see the state $\frac{1}{\limfunc{Tr}\left(
B\circ \left( A\circ \rho \right) \right) }\left( B\circ \left( A\circ \rho
\right) \right) $.

Alternatively, we may send the system in its state $\rho $ through the
compound filter $A\circ B$ which performs first $A$ then $B$ and measure the
resulting state at once. We then would get $\frac{1}{\limfunc{Tr}\left(
\left( A\circ B\right) \circ \rho \right) }\left( \left( A\circ B\right)
\circ \rho \right) $. In general, these two experiments lead to different
states. However, the normalizations are the same: 
\begin{eqnarray*}
\limfunc{Tr}\left( \left( A\circ B\right) \circ \rho \right) &=&\limfunc{Tr}
\left( \rho \left( \left( A\circ B\right) \circ I\right) \right) \text{ by
Condition (\ref{Duality}),} \\
&=&\limfunc{Tr}\left( \rho \left( A\circ B\right) \right) \text{ by
Condition (\ref{Unital}),} \\
&=&\limfunc{Tr}\left( \left( A\circ \rho \right) B\right) \text{ by
Condition (\ref{Duality}),} \\
&=&\limfunc{Tr}\left( \left( B\circ \left( A\circ \rho \right) \right)
\right) \text{ by Equality (\ref{halfDuality}).}
\end{eqnarray*}
Let us now assume that $A$ and $B$ commute. We have seen already that we
expect $AB=A\circ B$ and thus the compound filter has no "internal side
effects". Thus it should make no difference which of the two experiments we
conduct: we ought to obtain the same output state from the input $\rho $.
Hence, we obtain that for all states $\rho $, if $A$ and $B$ commute then: 
\begin{equation}
B\circ \left( A\circ \rho \right) =\left( A\circ B\right) \circ \rho \text{.}
\label{halfAsso}
\end{equation}

Using our duality assumption, we can deduce that for any effect $C\in 
\mathcal{E}\left( \mathcal{H}\right) $ we have, by successive applications
of Condition (\ref{Duality}): 
\begin{eqnarray*}
\limfunc{Tr}\left( \rho \left( \left( A\circ B\right) \circ C\right) \right)
&=&\limfunc{Tr}\left( (\left( A\circ B\right) \circ \rho )C\right) \\
&=&\limfunc{Tr}\left( \left( B\circ \left( A\circ \rho \right) \right)
C\right) \text{ by Equality (\ref{halfAsso}),} \\
&=&\limfunc{Tr}\left( \left( A\circ \rho \right) \left( B\circ C\right)
\right) \\
&=&\limfunc{Tr}\left( \rho \left( A\circ \left( B\circ C\right) \right)
\right) \text{.}
\end{eqnarray*}
As this is valid for all $\rho $ we conclude that if $AB=BA$ then $\left(
A\circ B\right) \circ C=A\circ \left( B\circ C\right) $. In fact, we shall
only require a special case of this relation, together with the observation
that $A^{2}=A\circ A$. We thus state:

\begin{condition}
\label{Associativity}\emph{(Weak associativity) }A sequential product $\circ 
$ needs to satisfy the relation: 
\begin{equation*}
A\circ \left( A\circ B\right) =\left( A\circ A\right) \circ B=A^{2}\circ B
\end{equation*}
for all $A,B\in \mathcal{E}\left( \mathcal{H}\right) $.
\end{condition}

\bigskip We shall require two more properties of a sequential product. First
of all, we desire the sequential product to be continuous. We saw that any
sequential product will be convex in its second variable which, with a
little work and our other assumptions, will grant continuity in the second
variable automatically. However, we also wish some form of continuity on the
first variable. We state:

\begin{condition}
\label{Continuity}\emph{(Continuity) }Let $B\in \mathcal{E}\left( \mathcal{H}
\right) $ be given. Then $$A\in \mathcal{E}\left( \mathcal{H}\right) \mapsto
A\circ B$$ is continuous in the strong operator topology.
\end{condition}

\bigskip The last condition which we impose on any sequential product is
preservation of pure states (up to normalization). A vector state, or a pure
state, is a rank-one orthogonal projection. Thus, let $\rho $ be a pure
state. If $A\circ \rho \not=0$ for $A\in \mathcal{E}\left( \mathcal{H}
\right) $ then it is reasonable that the state $\rho |A=\frac{1}{\limfunc{Tr}
\left( \rho A\right) }\left( A\circ \rho \right) $ conditioned on observing 
$A$ should again be pure.

\begin{condition}
\label{Purity}\emph{(Purity)} Let $p$ be a rank one orthogonal projection.
Then for all $A\in \mathcal{E}\left( \mathcal{H}\right) $ the effect $A\circ
p$ is of rank $1$ or $0$.
\end{condition}

\section{The Characterization Theorem}

We define a sequential product on $\mathcal{E}\left( \mathcal{H}\right) $ by
incorporating Conditions 1-5 from the previous section:

\begin{definition}
A sequential product $\circ $ on $\mathcal{E}\left( \mathcal{H}\right) $ is
a binary operation on $\mathcal{E}\left( \mathcal{H}\right) $ satisfying
Conditions 1-5, namely: for all $A,B\in \mathcal{E}\left( \mathcal{H}\right) 
$:

\begin{enumerate}
\item For all $\rho \in \mathcal{D}\left( \mathcal{H}\right) $ we have: 
\begin{equation*}
\limfunc{Tr}\left( \left( A\circ \rho \right) B\right) =\limfunc{Tr}\left(
\rho \left( A\circ B\right) \right) \text{,}
\end{equation*}

\item We have $A\circ I=I\circ A=A$,

\item We have $A^{2}\circ B=A\circ \left( A\circ B\right) $,

\item The map $E\in \mathcal{E}\left( \mathcal{H}\right) \mapsto E\circ B$
is continuous in the strong topology,

\item If $P$ is a pure state then $\frac{1}{\limfunc{Tr}\left( A\circ
P\right) }\left( A\circ P\right) $ is a pure state whenever $A\circ \rho
\not=0$.
\end{enumerate}
\end{definition}

\bigskip We check trivially that:

\begin{proposition}
\label{Sufficient}The product defined by $A,B\in \mathcal{E}\left( \mathcal{
\ H }\right) \mapsto A^{\frac{1}{2}}BA^{\frac{1}{2}}$ is a sequential
product on $\mathcal{E}\left( \mathcal{H}\right) $.
\end{proposition}

\bigskip We shall now prove the converse of the Proposition (\ref{Sufficient}
). We shall use the following notations. The set of all trace-class
operators on $\mathcal{H}$ is denoted by $\mathcal{T}\left( \mathcal{H}
\right) $. The set of positive trace-class operators is denoted by $\mathcal{
\ T}^{+}\left( \mathcal{H}\right) $. An element $A\in \mathcal{T}^{+}\left( 
\mathcal{H}\right) $ is \emph{pure} if whenever $0\leq B\leq A$ there exists 
$\lambda \in \left[ 0,1\right] $ such that $B=\lambda A$. Clearly, pure
elements are of the form $\lambda P$ for $\lambda \in \left[ 0,1\right] $
and $P$ a rank-one projection. A linear map $T:\mathcal{T}\left( \mathcal{H}
\right) \longrightarrow \mathcal{T}\left( \mathcal{H}\right) $ is \emph{\
positive} when $T\left( \mathcal{T}^{+}\left( \mathcal{H}\right) \right)
\subseteq \mathcal{T}^{+}\left( \mathcal{H}\right) $ and is pure when $T(A)$
is pure for all pure elements $A$ of $\mathcal{T}\left( \mathcal{H}\right) $.

\bigskip We now have:

\begin{theorem}
A map $\circ :$ $\mathcal{E}\left( \mathcal{H}\right) \times \mathcal{E}
\left( \mathcal{H}\right) \longrightarrow \mathcal{E}\left( \mathcal{H}
\right) $ is a sequential product on $\mathcal{E}\left( \mathcal{H}\right) $
if and only if for all $A,B\in \mathcal{E}\left( \mathcal{H}\right) $ we
have $A\circ B=A^{\frac{1}{2}}BA^{\frac{1}{2}}$.
\end{theorem}

\begin{proof}
The sufficient condition is Proposition (\ref{Sufficient}). Let us now prove
that the condition is necessary as well.

Let $\circ $ be a map satisfying Conditions 1-5. For $A\in \mathcal{E}\left( 
\mathcal{H}\right) $, we set $\Phi _{A}:B\in \mathcal{E}\left( \mathcal{H}
\right) \mapsto A\circ B$. If $\rho \in \mathcal{D}\left( \mathcal{H}\right) 
$ then: 
\begin{equation*}
\limfunc{Tr}\left( A\circ \rho \right) =\limfunc{Tr}\left( \rho (A\circ
I)\right) =\limfunc{Tr}(\rho A)
\end{equation*}
so $\Phi _{A}(\rho )\in \mathcal{T}^{+}\left( \mathcal{H}\right) $. By Lemma
(\ref{Convexity2}), the map $\Phi _{A}$ is affine on the convex set $
\mathcal{E}\left( \mathcal{H}\right) .$ Since $\mathcal{E}\left( \mathcal{H}
\right) $ generates algebraically the vector space $\mathcal{B}\left( 
\mathcal{H}\right) $ of all bounded linear operators on $\mathcal{H}$, it
follows that $\Phi _{A}$ has a unique linear extension, which we also denote
by $\Phi _{A}$ , to $\mathcal{B}\left( \mathcal{H}\right) $. Since $\limfunc{
Tr}\left( \Phi _{A}\left( \rho \right) \right) =\limfunc{Tr}\left( \rho
A\right) \leq 1$ for all $\rho \in \mathcal{D}\left( \mathcal{H}\right) $ we
conclude that the restriction of $\Phi _{A}$ to $\mathcal{T}\left( \mathcal{
H }\right) $ is a pure positive linear map from $\mathcal{T}\left( \mathcal{H
} \right) $ to $\mathcal{T}\left( \mathcal{H}\right) $. It follows from \cite
[ Theorem 3.1]{Davis76} that $\Phi _{A}:\mathcal{T}\left( \mathcal{H}\right)
\longrightarrow \mathcal{T}\left( \mathcal{H}\right) $ has one of the
following forms:

\newcounter{Lcount}
\begin{list}{\roman{Lcount}) }{\usecounter{Lcount}}
  
\item There exists $C\in \mathcal{B}\left( \mathcal{H}\right) $ such that
for all $\rho \in \mathcal{T}\left( \mathcal{H}\right) $ we have $\Phi
_{A}(\rho )=C^{\ast }\rho C$,

\item There exists a bounded conjugate linear map $C$ on $\mathcal{H}$ such
that for all $\rho \in \mathcal{T}\left( \mathcal{H}\right) $ we have $\Phi
_{A}\left( \rho \right) =C^{\ast }\rho ^{\ast }C$,

\item There exists $B\in \mathcal{B}\left( \mathcal{H}\right) ^{+}$ and some
orthogonal projection $P_{\psi }$ on the span of some unit vector $\psi $
such that for all $\rho \in \mathcal{T}\left( \mathcal{H}\right) $ we have $
\Phi _{A}(\rho )=\limfunc{Tr}\left( \rho B\right) P_{\psi }$.

  \end{list}

We first deal with case (iii). In this case: 
\begin{equation}
\limfunc{Tr}\left( \omega \Phi _{A}(\rho )\right) =\limfunc{Tr}\left( \rho
B\right) \left\langle \omega \psi ,\psi \right\rangle  \label{Eq1}
\end{equation}
for all $\omega \in \mathcal{T}^{+}\left( \mathcal{H}\right) $. Let $\left(
\omega _{i}\right) _{i\in \Lambda }$ be an increasing net in $\mathcal{B}
\left( \mathcal{H}\right) $ which converges to $I$ in the strong operator
topology.

Applying (\ref{Eq1}) we have: 
\begin{eqnarray*}
\limfunc{Tr}\left( \rho A\right) &=&\limfunc{Tr}\left( \Phi _{A}\left( \rho
\right) \right) =\lim_{i\in \Lambda }\limfunc{Tr}\left( \omega _{i}\Phi
_{A}\left( \rho \right) \right) \\
&=&\limfunc{Tr}\left( \rho B\right) \lim_{i\in \Lambda }\left\langle \omega
_{i}\psi ,\psi \right\rangle =\limfunc{Tr}\left( \rho B\right)
\end{eqnarray*}
for every $\rho \in \mathcal{T}\left( \mathcal{H}\right) $. Hence $B=A$ and
we have: 
\begin{equation}
A\circ \rho =\limfunc{Tr}\left( \rho A\right) P_{\psi }  \label{Eq2}
\end{equation}
for all $\rho \in \mathcal{T}\left( \mathcal{H}\right) $. Applying (\ref{Eq2}
) and Condition (\ref{Duality}) we conclude that: 
\begin{eqnarray}
\limfunc{Tr}\left( \rho A\right) \left\langle \omega \psi ,\psi
\right\rangle &=&\limfunc{Tr}\left( \omega \left( A\circ \rho \right) \right)
\notag \\
&=&\limfunc{Tr}\left( \left( \omega \circ A\right) \rho \right)  \notag \\
&=&\limfunc{Tr}\left( \omega A\right) \left\langle \rho \psi ,\psi
\right\rangle  \label{Eq3}
\end{eqnarray}
for all $\rho ,\omega \in \mathcal{D}\left( \mathcal{H}\right) $. In (\ref
{Eq3}), let $\rho =P_{\varphi }$ for some unit vector $\varphi \in \mathcal{
H }$ with $\left\langle \psi ,\varphi \right\rangle =0$. Then $\limfunc{Tr}
\left( P_{\varphi }A\right) \left\langle \omega \psi ,\psi \right\rangle =0$
for all $\omega \in \mathcal{D}\left( \mathcal{H}\right) $. Hence, $
\left\langle A\varphi ,\varphi \right\rangle =0$ so $A\varphi =0$ since $
A\geq 0$. It follows that $A=\lambda P_{\psi }$ for some $\lambda \in \left[
0,1\right] $. By (\ref{Eq2}) we have: 
\begin{eqnarray}
A\circ \rho &=&\lambda \limfunc{Tr}\left( \rho P_{\psi }\right) P_{\psi
}=\lambda \left\langle \rho \psi ,\psi \right\rangle P_{\psi }  \notag \\
&=&\sqrt[2]{\lambda }P_{\psi }\rho \sqrt[2]{\lambda }P_{\psi }=A^{\frac{1}{2}
}\rho A^{\frac{1}{2}}\text{.}  \label{Eq4}
\end{eqnarray}
We now show that the map $B\mapsto A\circ B$ is normal. First, notice that
if $B,C\in \mathcal{E}\left( \mathcal{H}\right) $ and $B\leq C$ then $C-B\in 
\mathcal{E}\left( \mathcal{H}\right) $ and we have: 
\begin{equation*}
A\circ C=A\circ \left[ B+(C-B)\right] =A\circ B+A\circ \left( C-B\right)
\geq A\circ B\text{.}
\end{equation*}
Next suppose that $\left( B_{i}\right) _{i\in \Lambda }$ is an increasing
net converging to $B$ in the strong operator topology. By Condition (\ref
{Duality}) we have: 
\begin{eqnarray*}
\lim_{i\in \Lambda }\limfunc{Tr}\left( \rho \left( A\circ B_{i}\right)
\right) &=&\lim_{i\in \Lambda }\limfunc{Tr}\left( \left( A\circ \rho \right)
B_{i}\right) =\limfunc{Tr}\left( \left( A\circ \rho \right) B\right) \\
&=&\limfunc{Tr}\left( \rho \left( A\circ B\right) \right) \text{.}
\end{eqnarray*}
Hence $\left( A\circ B_{i}\right) _{i\in \Lambda }$ converges to $A\circ B$
in the ultraweak topology. Since $\left( A\circ B_{i}\right) _{i\in \Lambda
} $ is an increasing net, it converges strongly to $A\circ B$ \cite
{Davis76,Rudin91}. To complete case (iii), for $B\in \mathcal{E}\left( 
\mathcal{H}\right) $ there exists an increasing net $\left( \rho _{i}\right)
_{i\in \Lambda }$ in $\mathcal{D}\left( \mathcal{H}\right) $ converging
strongly to $B$. Applying (\ref{Eq4}) and the normality of $B\mapsto A\circ
B $ we conclude that: 
\begin{equation*}
A\circ B=\lim_{i\in \Lambda }A\circ \rho _{i}=\lim_{i\in \Lambda }A^{\frac{1 
}{2}}\rho _{i}A^{\frac{1}{2}}=A^{\frac{1}{2}}BA^{\frac{1}{2}}\text{.}
\end{equation*}

We now treat case (i) and omit case (ii) which is dealt with in a similar
manner as (i). By normality we therefore can assume that there exists $C\in 
\mathcal{B}\left( \mathcal{H}\right) $ such that for all $B\in \mathcal{E}
\left( \mathcal{H}\right) $ we have $A\circ B=C^{\ast }BC$. Now 
\begin{equation*}
A=A\circ I=C^{\ast }C=\left\vert C\right\vert ^{2}
\end{equation*}
so $\left\vert C\right\vert =A^{\frac{1}{2}}$. By the Polar Decomposition
Theorem, there exists a partial isometry $U$ on $\mathcal{H}$ such that $
C=UA^{\frac{1}{2}}$. Then 
\begin{equation}
A=A^{\frac{1}{2}}U^{\ast }UA^{\frac{1}{2}}\text{.}  \label{Eq5}
\end{equation}

We now assume that $A$ is invertible. It follows from (\ref{Eq5}) that $
U^{\ast }U=1$. Applying Condition (\ref{Duality}) gives: 
\begin{eqnarray*}
\limfunc{Tr}\left( A^{\frac{1}{2}}U^{\ast }BUA^{\frac{1}{2}}\rho \right) &=& 
\limfunc{Tr}\left( BA^{\frac{1}{2}}U^{\ast }\rho UA^{\frac{1}{2}}\right) \\
&=&\limfunc{Tr}\left( UA^{\frac{1}{2}}BA^{\frac{1}{2}}U^{\ast }\rho \right)
\end{eqnarray*}
for every $\rho \in \mathcal{D}\left( \mathcal{H}\right) $. It follows that: 
\begin{equation}
A^{\frac{1}{2}}U^{\ast }BUA^{\frac{1}{2}}=UA^{\frac{1}{2}}BA^{\frac{1}{2}
}U^{\ast }  \label{Eq6}
\end{equation}

for every $B\in \mathcal{E}\left( \mathcal{H}\right) $. In particular, with $
B=I$ we have $A=UAU^{\ast }$ and since $U^{\ast }U=I$ we have $UA=AU$. Hence 
$A=AUU^{\ast }$ and since $A$ is invertible, we have $UU^{\ast }=I$. It
follows that $U$ is unitary. Moreover, from (\ref{Eq6}) we have that: 
\begin{equation*}
A^{\frac{1}{2}}U^{\ast }BUA^{\frac{1}{2}}=A^{\frac{1}{2}}UBU^{\ast }A^{\frac{
1}{2}}
\end{equation*}
and using the invertibility of $A$ again, we obtain that $U^{\ast
}BU=UBU^{\ast }$ for every $B\in \mathcal{E}\left( \mathcal{H}\right) $. It
follows that $U^{2}B=BU^{2}$ for all $B\in \mathcal{E}\left( \mathcal{H}
\right) $ so $U^{2}=\mu I$ with $\mu \in \mathbb{C}$ such that $\left\vert
\mu \right\vert =1$.

We now apply Condition (\ref{Associativity}) and obtain: 
\begin{eqnarray*}
A^{2}\circ B &=&A\circ (A\circ B)=A\circ \left( A^{\frac{1}{2}}U^{\ast
}BUA^{ \frac{1}{2}}\right) \\
&=&A^{\frac{1}{2}}U^{\ast }A^{\frac{1}{2}}U^{\ast }BUA^{\frac{1}{2}}UA^{ 
\frac{1}{2}} \\
&=&U^{\ast 2}ABAU^{2}=ABA\text{.}
\end{eqnarray*}
Replacing $A$ by $A^{\frac{1}{2}}$ we thus get $A\circ B=A^{\frac{1}{2}}BA^{ 
\frac{1}{2}}$ for all $B\in \mathcal{E}\left( \mathcal{H}\right) $.

Now, let $A\in \mathcal{E}\left( \mathcal{H}\right) $ not invertible.\ Then
for all $i\in \mathbb{N}\backslash \left\{ 0\right\} $ we set $A_{i}=\left(
1+\frac{1}{i}\right) ^{-1}\left( A+\frac{1}{i}I\right) $ and note that $
A_{i}\in \mathcal{E}\left( \mathcal{H}\right) $ is invertible. The sequence $
\left( A_{i}\right) _{i\in \mathbb{N}\backslash \{0\}}$ converges strongly
to $A$. It follows from Condition\ (\ref{Continuity}) that $A\circ B=A^{ 
\frac{1}{2}}BA^{\frac{1}{2}}$ for all $B\in \mathcal{E}\left( \mathcal{H}
\right) $.
\end{proof}

\providecommand{\bysame}{\leavevmode\hbox to3em{\hrulefill}\thinspace} 
\providecommand{\MR}{\relax\ifhmode\unskip\space\fi MR } 
\providecommand{\MRhref}[2]{  \href{http://www.ams.org/mathscinet-getitem?mr=#1}{#2}
} \providecommand{\href}[2]{#2}

\end{document}